 \documentclass[twocolumn,trackchanges]{aastex631}

\usepackage{graphicx}

\begin{document}


\title{Observational Astrophysics Research through Cross Institutional Research Partnerships}
\author{Kristen C. Dage} \thanks{NASA Einstein Fellow}
\affiliation{International Centre for Radio Astronomy Research $-$ Curtin University, GPO Box U1987, Perth, WA 6845, Australia}
\affiliation{Dept. Physics \& Astronomy, Wayne State University,  666 W. Hancock St, Detroit, MI, 48201, USA}%

  \email{kristen.dage@curtin.edu.au}

  \author{Linda Strubbe}
\affiliation{Strubbe Educational Consulting, Vancouver, British Columbia, Canada}

\author{Rhianna Taub}%
\affiliation{Dept. Physics \& Astronomy, Wayne State University,  666 W. Hancock St, Detroit, MI, 48201, USA}%

\author{Amna Khalyleh}
\affiliation{Henry Ford College, 5101 Evergreen Rd, Dearborn, MI 48128, USA \\}
\author{Malu Sudha}%
\affiliation{Dept. Physics \& Astronomy, Wayne State University,  666 W. Hancock St, Detroit, MI, 48201, USA}%

\author{Teresa Panurach}
\affiliation{Norfolk State University, 700 Park Avenue, Norfolk VA 23504, USA}
\author{Muhammad Ridha Aldhalemi}
\affiliation{Henry Ford College, 5101 Evergreen Rd, Dearborn, MI 48128, USA \\} 

\author{Zainab Bustani}
\affiliation{Henry Ford College, 5101 Evergreen Rd, Dearborn, MI 48128, USA \\}

\author{Dominic DeYonker}
\affiliation{Henry Ford College, 5101 Evergreen Rd, Dearborn, MI 48128, USA }
\author{Mariam Ismail Fawaz}
\affiliation{Henry Ford College, 5101 Evergreen Rd, Dearborn, MI 48128, USA \\}

\author{Hans J. Harff}
\affiliation{Henry Ford College, 5101 Evergreen Rd, Dearborn, MI 48128, USA \\}

\author{Timothy McBride}
\affiliation{Henry Ford College, 5101 Evergreen Rd, Dearborn, MI 48128, USA \\}

\author{Jesse Mason}
\affiliation{Henry Ford College, 5101 Evergreen Rd, Dearborn, MI 48128, USA \\}

\author{Anthony Preston}
\affiliation{Henry Ford College, 5101 Evergreen Rd, Dearborn, MI 48128, USA \\}
\affiliation{University of Michigan, 1085 S.\ University Ave. Ann Arbor, MI 48109 USA\\}

\author{Cortney Rinehart}
\affiliation{Henry Ford College, 5101 Evergreen Rd, Dearborn, MI 48128, USA \\}

\author{Ethan Vinson}
\affiliation{Henry Ford College, 5101 Evergreen Rd, Dearborn, MI 48128, USA \\} 

\author{Vanessa Wilson}
\affiliation{University of Michigan-Dearborn, 4901 Evergreen Rd, Dearborn, MI 48128, USA}%
\author{Edward M. Cackett}
\affiliation{Dept. Physics \& Astronomy, Wayne State University,  666 W. Hancock St, Detroit, MI, 48201, USA}%
\author{William I. Clarkson}
\affiliation{University of Michigan-Dearborn, 4901 Evergreen Rd, Dearborn, MI 48128, USA}%

\section{\label{sec:level1}Motivation}
 Undergraduate research in STEM can be a transformative experience, especially for Community College students, many of whom come from under-represented backgrounds. However, undergraduate research at Community Colleges is relatively rare, due in part to their focus on teaching excellence \citep{hewlett18}. Research partnerships with four year colleges about exciting and accessible science topics, facilitated by researchers such as postdoctoral fellows, can be a powerful path forward. We describe our program, the Dead Stars Society (DSS), as a successful example of such a research partnership, focused on the topic of observational astronomy.
 
Community colleges (CCs) are a  crucial and often overlooked institution in the goal of improving equity, diversity, and inclusion in STEM, and physics / astronomy in particular \footnote{\url{https://www.aps.org/apsnews/2024/02/physics-needs-community-colleges}}. In the United States, CCs are more diverse than four year colleges (4YCs) on many axes, including race/ethnicity, age, students who are first-generation in college, students who have dependent children, veterans, and students with disabilities \citep{NAP24622,doi:10.1177/00915521231163903}. As of 2020 just over 50\% of US college students were enrolled at a CC \citep{2024PhTea..62g.620S}.%

Participating in undergraduate research has many benefits (\cite{NAP24622, https://doi.org/10.1002/tea.21341, doi:10.1177/00915521211026682,hewlett18,higgins} and also the EP3 guide \cite{ep3}): 1) It helps students learn what STEM is and can encourage them to enter the field.  2) It helps to build STEM self-efficacy  and promotes STEM identity (particularly important for students from under-represented backgrounds). 3) It can be hugely beneficial for CC students to successfully transfer to a 4YC. 4) It helps students to complete their STEM degrees and reduce time to degree completion. 5) It builds interest in attending graduate school, and it helps them navigate the process of applying to graduate school.

The mission of CCs is heavily focused on providing education through teaching excellence, and there is less emphasis and financial support for undergraduate research culture \citep{hewlett16}. CC instructors may not have the time or incentive to lead student research programs on top of their teaching commitments. Engagement with external researchers at 4YCs is also not common, as there is often an unspoken stigma about attending a CC (\cite{hewlett18} and the personal experience of authors of this article).

However, STEM research at CCs is possible, and observational astronomy offers a powerful gateway. Astronomy is exciting, and a science about something almost everyone has access to, the sky. It is a very visual science, which means that the physics concepts can be connected to something accessible to students, regardless of their math levels. We are now entering the era of big survey telescopes, and astronomy is becoming data driven. This means that students who engage in astronomy research are learning transferrable coding and data visualization skills that are relevant to a whole host of careers outside academia. No specialized equipment is needed, and all of the data are publicly available.
We offer the DSS as a successful example of a CC-focused research partnership that  has features that we believe other institutions can try as well. This program was started by a professional astronomer who, as a former CC student, was personally motivated to provide research opportunities and professional development for CC students interested in STEM careers. The main framework of this program was a collaborative process between the professional researchers, the CC instructor and the students, with advice from the 4YC faculty. This is an ongoing partnership that began in January 2023, to conduct observational astronomy research, connecting students from a CC, Henry Ford College (HFC), and professional researchers at two 4YCs, University of Michigan-Dearborn (UM-D) and Wayne State University (WSU). HFC is located next door to UM-D in  Dearborn, Michigan, while WSU is about 10 miles away in Detroit. 

We started with one student in January 2023, focusing on X-ray astronomy, and gradually grew the number of students, while developing a research partnership structure that could be adapted to other contexts and environments. In early 2024, we switched focus from X-ray astronomy to preparing for science with the Vera C. Rubin Observatory. In 2025, the students started a project to do timing of neutron star low mass X-ray binaries to look for quasi periodic oscillations in data from NICER and NuSTAR \citep[e.g.,][]{2024RNAAS...8..243K}. %
A key feature of our project was that after learning data analysis techniques from the professional researcher, the student researchers developed their own research manuals, which include a brief description of the high level science case, and detailed instructions how to carry out the analysis. This both promotes ownership of learning and allows the project to scale to many students with relatively efficient investment of the professional researcher's time. Overall 12 students from the three institutions have been part of the DSS, with an active cohort of around 5 students at a given time.

\section{Research partnership structure}

The structure we developed was a collaborative process with equitable input from all members involved, especially including input from the student researchers.

\subsection{People}
The DSS is a partnership between postdoctoral professional researchers at WSU and a HFC physics/astronomy instructor, with high level support from 4YC faculty  from WSU and UM-D. The professional researchers (some of whom are former CC students) wanted to emphasize a focus on professional development and transferrable skills, supporting the student researchers while making progress on active questions in their own research areas, such that there was a good balance of effort for publishable scientific results for the professional researchers. After the initial program setup, this wound up being a relatively modest investment of time from the professional researchers (a few hours per week). 

We collaboratively agreed that the CC instructor's role would be supporting students while having high teaching effort, without getting overworked. The 4YC faculty, who were the most time constrained, were able to help with administrative issues, and with the longevity of the program. WSU and UM-D have large transfer student populations who often come from CCs, and the 4YC faculty were happy to support this as a pathway to support any  CC students who may want to transfer to their school. 

The cohort of student researchers were all compensated for their participation, either as an independent study or through paid opportunities  with their time commitments varying from a few hours per week up to 20, depending on their situation.

The CC instructor recruited student researchers, at first based on their interest/participation from physics or astronomy classroom interactions, and eventually through word-of-mouth. While there is a nominal GPA cutoff, students who demonstrated that their GPA was improving were also invited to participate. The student researchers were recruited for their curiosity, enthusiasm, and motivation. Their future career plans were not part of the consideration. Their identities reflect the diverse demographics of their community. \footnote{Appropriate student demographic data for these institutions are not readily available, but we note that many students at HFC and UM-D are of Middle Eastern descent / Arab American (often counted as ‘Caucasian’). }

\begin{figure}
    \centering
    \includegraphics[width=3.9in]{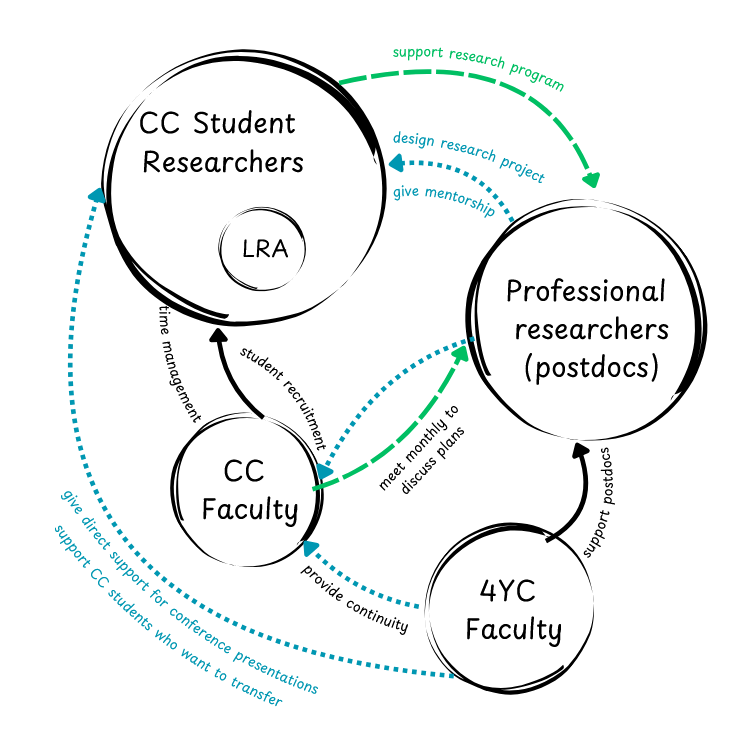}
    \caption{Cosmic web of participants. LRA refers to `lead research assistant'.  }
    \label{fig:enter-label}
\end{figure}
The CC instructor and other students selected a lead research undergraduate assistant (LRA), who is responsible for team paperwork, i.e., texting everyone to remember to submit their timecards),
for team task management, (i.e., keeping people on task),
and for team \textit{espirit de corps} (e.g., making a run to get shawarma or boba). 
Figure 1 shows the relationship between the people in this project. Some of the student researchers are also funded to do non-research related things, such as working with the planetarium, which the LRA also manages.
\subsection{Workflow}

For student-level research workflow, we co-developed a structure centered on peer instruction: The professional researcher teaches a data analysis technique to two or three student researchers (usually in person, but we have also done it over video conferencing). Those student researchers use the research technique on relevant data and then write a guide to doing the research, which also includes a high-level introduction to the science cases from the professional researchers. Then they use the guide to teach other student researchers. Holes in the guide are discovered and filled in by the team and, if needed, the professional researcher (via Slack). In summary, learn it, do it, write it, teach it, recycle it. This leverages the power of collaborative learning while ensuring that the entry point to the project $-$the analysis guides$-$ are accessible at the level of the student researchers, and critically, that the student researchers have ownership of their learning, autonomy, and engagement with the project. We emphasize that this structure was developed with the input and feedback from the student researchers.

\subsection{Research Projects}

In the last two years, we have embarked on two different science projects as a team. The first (Jan-Dec 2023) was reducing and analyzing X-ray data from the \textit{Chandra} X-ray Observatory to search for candidate intermediate mass black holes, and the second was interrogating new research tools developed for Rubin Observatory (Jan-Dec 2024). These projects were apt for the student researchers because once the student researchers overcame initial technical hurdles, they were able to become experts in a specific type of analysis and were able to analyze 150 data sets successfully.

The analysis they carried out needed to be done on scale, but could not be fully automated, and required human intervention. For example in project 1, the students reduced 250 GB of observations, and did a detailed study of 800 individual sources. The source coordinates and observations were listed in a spreadsheet, and the LRA assigned different sources and observations to different student researchers, allowing autonomy while still being supported by the professional researcher when they ran into trouble. 

The manuals are publicly available at a website maintained by the student researchers \footnote{\url{https://sites.google.com/view/deadstarssociety}}. Scientific results are being completed and written up by the professional researcher to be published in astronomy journals with students who contributed analysis included as co-authors \citep{2025ApJ...979...82D,2025arXiv250723332D}.

\subsection{Beyond the Science}

Beyond research engagement and transferrable skills, we also wanted to focus on professional development.  We were able to provide training in introductory python and data management techniques. 

The Rubin Observatory staff also provided a custom workshop via video conference on specific analysis notebooks developed by the observatory for scientists. We developed and delivered a workshop on ``Mentoring Relationships and Owning Your Career'', which covered advice on navigating the mentor-mentee relationship, and making a strategic plan.

The student researchers were also able to network at astronomy conferences, both local and national (a local conference on black holes, the Rubin Community Workshop and the 245th American Astronomical Society). The local Compact Objects in Michigan and Ontario conference (COMO) is a yearly one day conference to gather all of the professional researchers in the region who are interested in black hole and neutron star research. It is a relaxed conference with a focus on providing opportunities for early career researchers to present. Three of our student researchers attended in 2023, with one presenting on the X-ray research. In 2024, HFC hosted COMO on campus, attended by 5 student researchers, with one presenting a brief overview of both the \textit{Chandra} and Rubin based projects. 
The student presenters worked with the professional researcher to develop their talk, and practiced it multiple times in person or virtually with the faculty and professional researchers, and also with external astrophysicists with relevant expertise.

We initiate new student researchers into the Dead Star Society through a ceremony which involves a poetry reading (typically Invictus), and the bestowal of their “ASIB” (a star is born) token. Additional tokens are given out at anniversaries, and to commemorate achievements, like presentations and papers.
As part of their paid role, the student researchers were also tasked with various assignments around HFC assisting the CC instructor. This included working with the planetarium and engaging in public outreach at a number of events.  

\begin{table*}
\caption{Known barriers for CC students to engage in research identified in  Hewlett 2018 \cite{hewlett18}, and how the DSS addresses them.}
\label{table:tabls}
\begin{tabular}{ll}
Barrier from Hewlett 2018 \cite{hewlett18}                  & Our Solution                                                                               \\ \hline
Limited financial resources  & Students analyze publicly available data                                        \\
Incompatible faculty model & Students team up with external research mentors outside their CC                                                               \\
Limited student preparation  & Students develop research manuals aimed at entry level researchers\\
Isolation from networks      & CC faculty teams up with researchers at 4YC / observatories                                                            
\end{tabular}
\end{table*}

We hold weekly meetings that the students attend as their schedule allows, and the HFC instructor keeps a physical task table 
and has the student researchers check off their progress for that week. The HFC instructor and professional researchers meet on a $\sim$ monthly basis, to discuss projects,  looking about 4-6 months ahead. 

\section{Outcomes, Benefits and Challenges}

Several of the students have  transferred to 4YCs to complete STEM Bachelor's degrees. 

The professional researchers also heavily benefited from this partnership, both personally and professionally. Two major outcomes included publications and mentorship experience. 

For the wider astronomical community, we have made the manuals developed by the student researchers publicly available, to serve as a resource for other undergraduate students who may wish to do observational astrophysics research. HFC also hosted their first ever astronomy conference.

One of the first major challenges we encountered was computers and software installation. Our workaround was for the students to use virtual machines on their personal laptops, until the LSST Discovery Alliance was able to buy us inexpensive, refurbished laptops.
Communication was another challenge. While the institutions are geographically close, the professional researchers' primary affiliation was not with HFC, with one researcher moving institutions in 2024, and it was not always possible to commit to spending time at HFC. It was important to spend some initial time face to face to establish a relationship with the student researchers, and for analysis method training,  until we knew what issues the student researchers were likely to run into. It was also important to build up a relationship/rapport where student researchers felt comfortable asking questions. In terms of analysis troubleshooting, Slack was extremely helpful. We created a Slack channel for help with the particular tasks, and then student researchers could upload the pictures where the professional researcher could answer quickly whether there was a quick solution, or if the student should just move on to the next object. 

We note that because the student researchers are very early on in their careers, with not all of them having taken an introductory astronomy class, they were less aware of the jargon, and terminology professional astronomers might take for granted. The student researchers were very willing, and not shy at all, to express if further clarification was required, which significantly contributed to the success.

\section{Conclusions \& Recommendations}
 We strongly encourage cross institutional collaboration between professional researchers and under-resourced institutions; it is a symbiotic relationship that allows students to engage in astronomy research, facilitated by CC staff while furthering the research goals and providing mentoring opportunities to early career scientists. CC students are passionate and highly motivated. We believe that engagement in astronomy research is useful to undergraduate students, 
and can teach them transferrable skills useful throughout their future careers, astronomy or otherwise. Our hope is that this research partnership structure will be useful to other professional researchers and students at institutions without direct access to astronomy research opportunities. We feel that many astronomical observatories could also benefit from partnerships with CCs. 

Our proposed program addresses many of the common barriers to integrating research into the CC student experience identified in \cite{hewlett18} (see Table 1) and addresses many recommendations in \citep{doi:10.1177/00915521211026682}, and  \citep{Savrda2024-fe}. 

\begin{itemize}
    \item \textbf{Creating a partnership:} We encourage \textbf{collaborative partnerships} between 4YCs and  under-resourced institutions facilitated by a professional researcher and CC instructor, with involvement from the 4YC faculty for longevity. We agree with  \cite{Savrda2024-fe}, who also emphasize the importance of communication and equitable collaboration with CC faculty and their essential role in these partnerships.
    The design of the partnership is an \textbf{iterative process} between all parties. This promotes \textbf{equality, agency and partnership status}. \textbf{Research is a job.} There may be flexibility or pockets of money to fund students, but be aware that each institution has its own unique set of funding opportunities and constraint, and the details of how will be very specific to each institution. We note that multi-institutional proposing teams can often make a compelling case for large grants.

    \item \textbf{Project design:} Design a project with a relatively directed science goal; it is important for novice researchers to have \textbf{defined parameter space} with room for agency. We found that having a project that, once the students mastered the analysis technique, they could deploy on a large source list, was highly beneficial because it got the research done more efficiently, and promoted\textbf{ ownership of learning} for the student researchers. 
    \item \textbf{How students learned and worked:}  It is important to have the students work together as a\textbf{ cohort}.  Learning through teaching is valuable, promotes ownership of learning, and helps make the topics accessible to students who haven’t taken an astronomy course before (in some cases). Emphasize the importance of a growth mindset by reiterating that research takes time and is often about\textbf{ learning from failure}. It is important to get the student researchers involved in astronomical community through talks and external collaboration. With an already \textbf{existing structure}, this research could be carried out virtually. However, it is very valuable to have person-to-person time at the beginning to get it started. 
  
   \item \textbf{Program evaluation is important.} In additional to informally checking in with students periodically, we also had an external evaluator collect feedback via a focus group. We recommend in the future a\textbf{ robust measurement of the impact} on students' science identities, ownership of learning, self efficacy and sense of belonging, as well as feedback from professional researchers, and CC and 4YC faculty, about what they get out of the process/project.  We recommend engaging an external evaluator to design a robust evaluation, which could include, for example, validated surveys and focus groups. 

\end{itemize}

\section{Acknowledgements}
We acknowledge support from LSST Discovery Alliance through funding provided by the Heising-Simons Foundation grant \#2020-1916.
\bibliographystyle{aasjournal}
\bibliography{aipsamp}

\end{document}